# Damage Preserving Transformation for Materials with Microstructure

Philip P. Müller[a], Falk K. Wittel[a], David S. Kammer[a,*]

[a]*Institute for Building Materials (IfB), ETH Zuerich, Laura-Hezner-Weg 7, 8093, Zuerich, Switzerland*

**Abstract**

The failure of heterogeneous materials with microstructures is a complex process of damage nucleation, growth and localisation. This process spans multiple length scales and is challenging to simulate numerically due to its high computational cost. One option is to use domain decomposed multi-scale methods with dynamical refinement. If needed, these methods refine coarse regions into a fine-scale representation to explicitly model the damage in the microstructure. However, damage evolution is commonly restricted to fine-scale regions only. Thus, they are unable to capture the full complexity and breath of the degradation process in the material. In this contribution, a generic procedure that allows to account for damage in all representations is proposed. The approach combines a specially designed orthotropic damage law, with a scheme to generate pre-damaged fine-scale microstructures. Results indicate that the damage approximation for the coarse representation works well. Furthermore, the generated fine-scale damage patterns are overall consistent with explicitly simulated damage patterns. Minor discrepancies occur in the generation but subsequently vanish when explicit damage evolution continuous; for instance under increased load. The presented approach provides a methodological basis for adaptive multi-scale simulation schemes with consistent damage evolution.

*Corresponding author
Email addresses: `phimuell@ethz.ch` (Philip P. Müller), `fwittel@ethz.ch` (Falk K. Wittel), `dkammer@ethz.ch` (David S. Kammer)

**Contents**

# 1. Introduction

At a certain scale even heterogeneous materials will appear homogeneous and some can even be considered isotropic. Among others, this is true for concrete, one of the most widely used commodity on earth, a mixture made of sand, aggregates, cement, water and chemical admixtures. The growth of damage inside concrete is highly affected by the particular microstructure, where, depending on the scale, aggregates or even sand grains act either as focal points for stresses or obstacles for damage.

Damage initiates at very small scales, long before the macroscopic structure itself will fail or crack. Instead, the damage leads to a reduction of the material's stiffness. Nevertheless, at one point the accumulated damage becomes so widespread, that even its smallest increase, will trigger the previously isolated nuclei to merge. This leads to a cascade of increasingly larger defects, culminating in the emergence of a macroscopic crack.

Continuum based methods are the methods of choice if large structures should be simulated, due to their computational efficiency. For taking into account intrinsic degenerative processes constitutive laws are used. One of the earliest, but still widely used laws for modelling damage in concrete was proposed by Mazars (Lemaître, 2001; Mazars et al., 1985). It employs a scalar damage variable to degrade the material's stiffness. However, even if the material was initially isotropic, damage will induce anisotropy into the material's behaviour. Clearly any scalar damage variable is inherently unable to capture this. During the years, a variety of anisotropic damage models were proposed to address this issue (Brancherie et al., 2009; Braun et al., 2021; H. Chen et al., 2016; Delaplace et al., 2008; Desmorat et al., 2007; Gaede et al., 2013). All of them consider the accumulated effects of the damage's growth, represented by internal state variables at the material points and by that disregard the actual microstructure, whose degeneration is the actual cause for the emerging damage.

To overcome this deficiency, the entire microstructure could be explicitly represented and simulated. Unfortunately, even with today's fast computers, this is only possible for small sizes. A way to overcome this barrier are multi-scale methods. They allow to invest computational power exactly where it is needed, by combining different representations. Although, many different methods were proposed over the past years, they can be classified to be either of hierarchical or of concurrent nature (Budarapu et al., 2017; Liu, 2018; Lloberas-Valls et al., 2012a; Matous et al., 2017; Unger and Eckardt, 2011).

Hierarchical methods, such as (Eliá et al., 2022; Rezakhani et al., 2017; X. Sun et al., 2019; Xu et al., 2023), are characterised by a full separation of scales, which allows to treat every level independently from each other. Thus, the information is passed between the different levels as one serves as input for the hierarchically higher level.

Opposed to this, concurrent methods, such as (Lloberas-Valls et al., 2012b; Miller et al., 2009; Xiao et al., 2004), lack the full separation of scales and typically decompose the computational domain into different regions. Imagine a typical setting where high accuracy is only needed inside a small part of the computational domain, for example around a crack tip. Ideally, one limits methods with high accuracy but large computational burden to these small regions, while the remaining part of the computational domain is described by much more efficient methods. The link between the different regions is established by a coupling scheme (Bitencourt et al., 2015; Farhat et al., 1991; Lloberas-Valls et al., 2012b; Unger and Eckardt, 2011), of which the Arlequin method is a particular general one (Anciaux et al., 2008; Bauman et al., 2008; Guidault et al., 2007; Unger and Eckardt, 2011; Wellmann et al., 2012).

Whenever the domain decomposition is not available in advance, one must resort to adaptive methods, which refine regions on demand (*e.g.*, P. Y. Chen et al., 2021; Evangelista, Alves, et al., 2020; Evangelista and Moreira, 2020; Rodrigues et al., 2018; Unger and Eckardt, 2011; Zhang et al., 2012). However, important questions are (i) how are the regions that need to be refined identified, and (ii) how is the loading history of the coarse connected to the initial state of the newly created fine scale representation? Especially (ii) does not seem to be addressed well in literature. Some authors assume that the coarse representation does not accumulate any damage prior to refinement, which is triggered by a stress or strain based criteria (L. Chen et al., 2021; P. Y. Chen et al., 2021; Rodrigues et al., 2018; Unger, Eckardt, and Konke, 2011). Other authors link the refinement criteria to a damage law (Lloberas-Valls et al., 2012b; Rezakhani et al., 2017; B. Sun et al., 2015; X. Sun et al., 2019; Xu et al., 2023). Since most damage laws employ a threshold below which it is inactive, refinement is triggered the first time the loading surpasses it. Both approaches restrict damage evolution only to refined regions, but have the advantage that the refined region is always undamaged at the beginning. Consequently, the entire load history of the coarse regions is disregarded and refinement might occur unnecessarily early, since even the smallest damage triggers a refinement.

In this paper we propose, to the best of our knowledge, a generic approach for the refinement step in adaptive concurrent multi-scale simulations, which allows to account for the damage evolution inside the coarse representation. To this end, we equip the coarse scale representation with its own anisotropic damage measure, which is based on a damage variable recently presented by Oliver-Leblond et al. (2021). Further, we develop a scheme to create fine scale representations with a certain initial damage pattern on the fly. This allows us to create fine scale regions that already contain an initial damage. Together, these two components will allow to include damage evolution in unrefined regions and to incorporate the coarse loading history upon refinement. In addition, the coarse damage measure will allows much better control of the refinement.

While our approach is generic and rather simple, its practical details highly depend on the selected represen-tations. Thus, we demonstrate it by applying it to one particular test case. The reminder of this paper is organised as follows: In Sec. 2, we explain our method in more detail and present the proposed techniques. In Sec. 3 we determine the parameters of our method and asses its applicability, before we draw final conclusions in Sec. 4.

## 2. Materials and Methods

The particular choice of the material's microstructure, also called motive, is in general arbitrary, but should follow principles of representative volume elements (RVE) (Lemaître and Desmorat, 2005). The state of a discrete representation with its inherent characteristic structure is fully given by $\mathfrak{r}$, that describes every single discrete element (right side of Fig. 1). In this representation, damage $\mathbb{D}(\mathfrak{r})$ is given by the irreversible degeneration of the constituting elements. On the left side of Fig. 1, the smeared continuum representation is shown, which lacks such an explicit microstructure and only considers cumulative effects of the damage through internal state variables added to the constitutive law. Here, damage is given by $\mathbf{D}$, which depends on the state $\vec{\kappa}$ at a particular location and is embedded in the constitutive law.

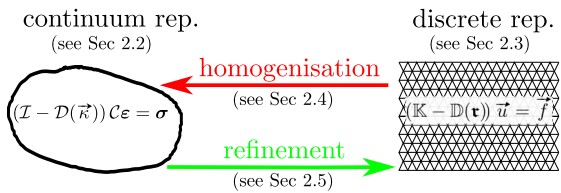

Figure 1: The continuum damage depends on the continuum state $\vec{\kappa}$ and has value $\mathcal{D}$. Formally $\mathcal{D}$ is a fourth order tensor, but in this work we are using the second order tensor $\mathbf{D}$ to represent damage. The discrete damage $\mathbb{D}(\mathfrak{r})$ depends on $\mathfrak{r}$ and hence on the state of all discrete elements of the lattice. The two representations as well as their damages are interconnected to each other by the homogenisation and refinement processes, respectively.
Scale of the lattice is exaggerated.

Since the continuum representation loses its validity once cracks localise, one must refine the continuum to its discrete twin in such way that all important aspects of the fracture will be captured accurately on the fine scale. The key for a meaningful adaptive modelling of the damage evolution lies in the transformation of the continuum to the discrete representation, that conserves the degraded mechanical behaviour found inside the continuum. One focus of this work is an approach to construct a discrete representation that respects the preceding damage present in the continuum representation.

Even though the procedure is generic and in principal not restricted to specific numerical material representations, this paper focuses on one particular choice. However, we will outline the generic way of working with the method (see Sec. 2.1), before we start with our specific choice. We exemplary chose a two-dimensional plane stress, isotropic material (see Sec. 2.2) with an underlying material heterogeneity represented by a triangular network of beam-truss elements with linear-elastic, brittle behaviour with quenched disorder of breaking thresholds (see Sec. 2.3). We then discuss the particular choice of the damage law as well as the reconstruction step (see Secs. 2.4, 2.5). To determine and test them, we use data obtained from numerical simulations (see Sec. 2.6).

### 2.1. Generic Damage Transforming Method

Initially, the domain is described as a continuum without any internal structure, whose state is fully described by the continuum state variable $\vec{\kappa}$. In the continuum, the damage evolution is fully govern by the damage function $\widehat{\mathbf{D}}(\vec{\kappa})$. Therefore, $\widehat{\mathbf{D}}(\vec{\kappa})$ can be interpreted as the macroscopic damage, that is expected for a hypothetical discrete representation with identical loading. Thus we can determine the function describing the macroscopic damage by homogenising the discrete damage $\mathbb{D}(\mathfrak{r})$. This leads to a perspective on the damage law that is different from the conventional one, where the damage law is calibrated against a physical material. Instead, here the law is calibrated against a particular numerical representation of the material.

When the continuum model experiences a certain damage limit, it is no longer suitable and has to be refined to a discrete representation. However, this discrete state has to be consistent with the previous continuum representation. This includes stiffness and damage, which have to be preserved as much as possible by the transformation. Determining this reconstruction process challenging, since it is by its very nature not unique.

### 2.2. Continuum Representation of 2D Isotropic Continua with Damage

To represent a two-dimensional isotropic material under plane stress, the Finite Element Method (FEM) and as damage measure continuum damage mechanics (CDM) is used (Lemaître, 2001; Lemaître, 1996; Lemaître and Desmorat, 2005). We use the well known material law:

$$\boldsymbol{\sigma} = (\mathcal{I} - \mathcal{D}) \, \mathcal{C} \boldsymbol{\varepsilon}, \qquad (1)$$

where $\boldsymbol{\sigma}$ and $\boldsymbol{\varepsilon}$ denote the continuum stress and strain tensors, respectively, $\mathcal{C}$ is the continuum stiffness tensor of the undamaged material, and $\mathcal{D}$ is the damage tensor. While in Eq. (1) damage is represented by a fourth order tensor, in this work we use the second order tensor $\mathbf{D}$ to describe damage (Lemaître and Desmorat, 2005, Sec. 1.1.4). This choice allows to account for anisotropic effects, while avoiding the complexity of fourth order tensors. Due to the choice of CDM, as continuum damage measure, the damage variable $\mathbf{D}$ can directly be identified with the damage function $\widehat{\mathbf{D}}(\vec{\kappa})$. Further, we identify $\vec{\kappa}$ as the continuum state variable given as

$$\vec{\kappa} := \begin{pmatrix} \kappa_x \\ \kappa_y \end{pmatrix}. \qquad (2)$$

A zoning approach is used to divide the principal strain space along an angle $\chi$, known as zone boundary, into an $x$- (shaded red parts) and $y$-zone (shaded yellow parts in Fig. 2). The two components $\kappa_x$ and $\kappa_y$ represent the maximal reached principal tensile strain in $x$ and $y$ direction, respectively, i.e.

$$\kappa_x := \max\left(\kappa_x, \langle\varepsilon_1\rangle_+\right), \qquad \kappa_y := \max\left(\kappa_y, \langle\varepsilon_2\rangle_+\right).$$

While $\varepsilon_1$ and $\varepsilon_2$ are the eigenvalues of the strain tensor $\boldsymbol{\varepsilon}$, its eigenvectors form a Givens rotation matrix of angle $\Gamma$, which is sometimes called eigenangle. The angle $\Gamma$, together with the boundary $\chi$, determines which zone the eigenvalues are associated with, see Fig. 2.

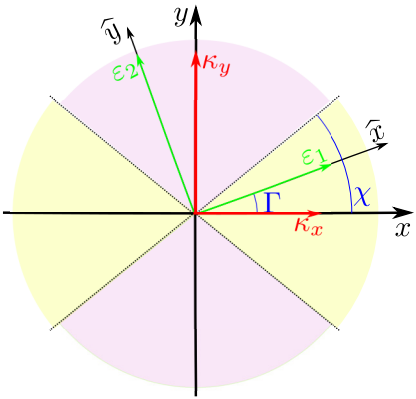

Figure 2: Interpretation of the zone boundary parameter $\chi$. While $\varepsilon_1$ and $\varepsilon_2$ are the eigenvalues of $\boldsymbol{\varepsilon}$, its eigenvectors are described by the value $\Gamma$. The eigenangle $\Gamma$ and the zone boundary $\chi$ determines which eigenvalue acts in which direction.

In the initial phase, where the CDM is used, the damage is still small. This allows us to make some simplifying assumptions on the damage variable $\mathbf{D}$, that will also affect $\widehat{\mathbf{D}}(\vec{\kappa})$: (i) We assume that the damage is orthotropic which reduces $\mathbf{D}$ to a diagonal matrix. For that reason, we will only consider the eigenvalues of $\mathbf{D}$, cf. Eq. (11). (ii) We assume that there is no coupling between the directions. Thus, a change of $\kappa_x$ ($\kappa_y$) will only affect the damage along the $x$-direction ($y$-direction). Later we will slightly relax this assumption, cf. Eq. (12).

### 2.3. Exemplary Material Motive

The example material motive chosen here is based on models proposed in Refs. (Herrmann et al., 1989; Mier, 2017), namely a regular triangular lattice but formed by 3$^{\mathrm{rd}}$ order Reddy truss-beam elements with characteristic lattice size $\ell$ (Reddy, 1997; Reddy et al., 1997). Using beams allows to include bending properties and the resulting lattice is able to represent a Cosserat continuum (Ostoja-Starzewski, 2008; Vardoulakis, 2019). The microscopical beams consist of an isotropic material with Young's modulus $E_b$ and Poisson's ratio $\nu_b$. A list of all used material parameters is given in Tab. I. In a multiscale simulation, $E_b$ has to be chosen such that the resulting behaviour of the discrete structure matches the one of the continuum, i.e. its stiffness tensor $\mathcal{C}$. However, since this paper studies the refinement step in isolation, without having an actual continuum phase, the choice of $E_b$ is actually irrelevant.

Table I: Parameters of the discrete material motive.

| Property | Value | Unit |
|---|---|---|
| $N_x$, $N_y$ | 300, 346 | $[-]$ |
| $L_x$, $L_y$ | 2, 1.998 | m |
| $H$ | 1 | m |
| $E_b$ | $1 \times 10^6$ | Pa |
| $\nu_b$ | 0.3 | $[-]$ |
| $k_\varepsilon$ | 3 | $[-]$ |
| $\lambda_\varepsilon$ | 0.02 | $[-]$ |
| $k_\Phi$ | 3 | $°$ |
| $\lambda_\Phi$ | 0.02 | $[-]$ |

*Lattice Geometry.* The motive is defined by the number of nodes ($N_x$, $N_y$) in $x$- and $y$-direction, the spatial extension in $x$-direction $L_x$, with resulting characteristic lattice size $\ell := L_x/(N_x - 1)$ and spatial $y$-extension $L_y := N_y \ell \sqrt{3}/2$. The influence of $\ell$ on the damage evolution is small (see Appendix B). An out-of-plane height of $H$ is assumed. To remove the symmetries of the lattice, topological disorder is introduced (Moukarzel et al., 1992; Wittel, 2006) by adding the random displacement

$$\vec{x}_i^\Delta := a \frac{\ell}{2} \vec{x}_i^* \tag{3}$$

to every internal node of the grid, where $\vec{x}_i^*$ is a random vector sampled uniformly from the unit circle (see Fig. 3a). The distortion is controlled by parameter $a \in [0, 1[$, known as distortion level.

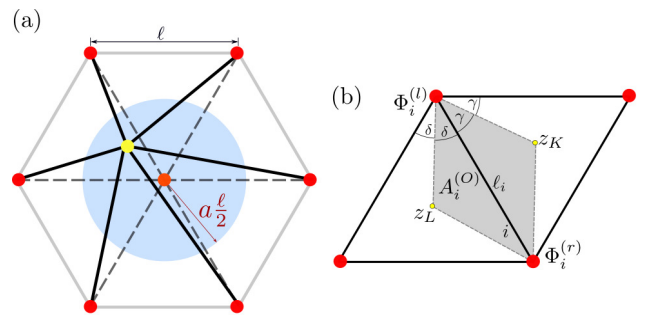

Figure 3: (a) Distortion of the central node, ignoring the distortion of the surrounding nodes. The location of the distorted node (yellow circle), is randomly selected within the blue circle of radius $a\ell/2$. Afterwards, the length of the beams are adjusted to match the new node location (black lines). (b) The thickness of beam $i$ is given as $t_i := A_i^{(O)}/\ell_i$, where $\ell_i$ is its length and $A_i^{(O)}$ is the area the beam is representing. Points $z_L$ and $z_K$ are centres of the adjacent triangles' incircles.

*Geometrical Properties of Beam-Truss Elements.* The thickness of beam $i$, denoted as $t_i$, depends on the lattice's geometry. It is given as $t_i := A_i^{(O)}/\ell_i$, where $A_i^{(O)}$ is the area represented by the beam and $\ell_i$ its length, see

Fig. 3b. The area $A_i^{(O)}$ is formally defined as the set of points that are closer to beam $i$ than any other beam, but are inside the lattice. It can be determined by finding the intersection of the angle's bisectors, i.e. centre of the incircle, of the two adjacent triangles denoted as $z_K$ and $z_L$ in Fig. 3b. In case the beam is part of the boundary $A_i^{(O)}$ is artificially doubled. This ensures that in a regular lattice all beams have the same axial rigidity.

*Damage Criterion Applied to the Beam-Truss Lattice.* In the discrete representation, damage is the irreversible failure of elements, namely the reduction of their contributing stiffness to an insignificant level. To determine if a beam has surpassed its loading capacity, the elliptical criterion

$$\left(\frac{\varepsilon_i}{\varepsilon_{i;th}}\right)^2 + \frac{\max\left(\left|\Phi_i^{(r)}\right|, \left|\Phi_i^{(l)}\right|\right)}{\Phi_{i;th}} =: \Psi_i \geq 1 \quad (4)$$

is used, where $\varepsilon_{i;th}$ and $\Phi_{i;th}$ are the beam's elongation and bending thresholds, respectively (Herrmann et al., 1989). Both thresholds are sampled independently from the Weibull distributions $\varepsilon_{i;th} \overset{\text{iid}}{\sim} \text{Weib}(k_\varepsilon, \lambda_\varepsilon)$ and $\Phi_{i;th} \overset{\text{iid}}{\sim} \text{Weib}(k_\Phi, \lambda_\Phi)$.

*The Discrete State Variable $\vec{r}$.* The discrete state is uniquely described by $\mathfrak{r}$. However, for the context of this paper the surrogate discrete state variable

$$\vec{r} := \begin{pmatrix} r_x \\ r_y \end{pmatrix} \quad (5)$$

is introduced and termed "discrete state variable". Since $\vec{r}$ has only two components it does not uniquely describe the damaged state. This ambiguity will be resolved by the reconstruction process (see Sec. 2.5). $\vec{r}$ is a purely mathematical quantity designed to have certain properties. First, its 1-norm $\widetilde{r} := \|\vec{r}\|_1 := |r_x| + |r_x|$ equals to $N_f/N_T$, where $N_f$ is the number of failed beams and $N_T$ the total number of beams in the lattice. $\widetilde{r}$ is also called the ratio of failed beams (rfb). Second, its components are defined by associating them to the $x$- and $y$-zone, respectively, similar to $\vec{\kappa}$ (see Sec. 2.2). But while $\kappa_x$ is connected to strains in the $x$-zone, $r_x$ is related to the amount of beams that have failed due to $\kappa_x$.

### 2.4. Determining the Damage Law for the Continuum

The damage function $\widehat{\mathbf{D}}(\vec{\kappa})$ will take the role of the damage variable $\mathbf{D}$ inside the constitutive equation (1). Thus, $\widehat{\mathbf{D}}(\vec{\kappa})$ has to be designed such that its evolution mimics the expected behaviour of $\mathbf{D}$ (see Sec. 3.2). For the extraction, which involves two steps, the `UniformSim` simulation data of fully discrete lattices is used (see Sec. 2.6.2).

*Step 1: Effective Material Stiffness Tensor $\mathcal{C}$.* First, the effective stiffness tensor $\mathcal{C}$ is calculated by homogenisation. After the convergence of each loading step, the following seven strain-states

$$\left\{\begin{pmatrix}1\\2\\3\end{pmatrix}, \begin{pmatrix}4\\5\\0\end{pmatrix}, \begin{pmatrix}6\\0\\0\end{pmatrix}, \begin{pmatrix}0\\7\\0\end{pmatrix}, \begin{pmatrix}0\\0\\8\end{pmatrix}, \begin{pmatrix}9\\0\\10\end{pmatrix}, \begin{pmatrix}0\\9\\8\end{pmatrix}\right\}, \quad (6)$$

denoted as $(\varepsilon_{xx}, \varepsilon_{yy}, 2\varepsilon_{xy})^{\mathrm{T}} \times 10^{-3}$, were applied to the lattice, while blocking further damage to measure the resulting stresses. This results in an overdetermined system of 21 equations for the 6 unknown coefficients of $\mathcal{C}$, which is solved by a least-square approach.

*Step 2: Determining the Damage Variable $\mathbf{D}$.* Second, the damage variable $\mathbf{D}$ is extracted from the effective stiffness tensors of the lattice. For this, a technique originally presented by Oliver-Leblond et al. (2021) is used. For completeness, the relevant equations are replicated to be

$$\mathbf{d}(\mathcal{T}) := \text{tr}_{1,2}[\mathcal{T}] = \mathcal{T}_{kkij}, \quad (7a)$$

$$K := \frac{1}{4}\text{tr}\left[\mathbf{d}(\mathcal{C})\right], \quad (7b)$$

$$\mathbf{D} := \mathbf{D}\left[\mathcal{C}, \widehat{\mathcal{C}}\right] := \frac{1}{2K}\left(\mathbf{d}(\mathcal{C}) - \mathbf{d}(\widehat{\mathcal{C}})\right). \quad (7c)$$

The tensor defined by Eq. (7a) is also known as dilatation second order tensor, while scalar $K$ of Eq. (7b) is the bulk modulus. Eq. (7c) combines the effective $\mathcal{C}$ and undamaged stiffness tensor $\widehat{\mathcal{C}}$ to the damage variable $\mathbf{D}$. $\mathbf{D}$ is by construction a real symmetric $2 \times 2$ matrix, thus fully characterised by its two eigenvalues $d^{(x)}$ and $d^{(y)}$ as well as a single scalar $\Gamma$, describing the rotation of its eigenbasis (see Fig. 2).

### 2.5. Process for the Construction of a Damaged Lattice

The reconstruction process, i.e. the creation of a discrete lattice with a particular damage, involves two components: (i) The transfer function $\widehat{\vec{r}}(\vec{\kappa})$, which transforms the continuum state $\vec{\kappa}$ into the discrete surrogate state variable $\vec{r}$. (ii) A scheme which transforms the surrogate state $\vec{r}$ into the full discrete state $\mathfrak{r}$. Hence, the scheme must be able to resolve the inherently present ambiguity in $\vec{r}$. As direct consequence of the definition of the discrete state $\vec{r}$ (see Eq. (5)), the transfer function is given as

$$\widehat{\vec{r}}(\vec{\kappa}) := \begin{pmatrix} \widehat{r}_x(\kappa_x) \\ \widehat{r}_y(\kappa_y) \end{pmatrix}. \quad (8)$$

As for the damage function $\widehat{\mathbf{D}}(\vec{\kappa})$, we are using data obtained from the `UniformSim` simulations (see Sec. 2.6.2) to empirically determine the function $\widehat{\vec{r}}(\vec{\kappa})$ that approximates $\vec{r}$. Due to the nature of $\vec{r}$ it is impossible to measure its components and thus to fit them directly. However, it is easy to measure and fit the quantity $\widetilde{r} := \|\vec{r}\|_1$.

Because of the specific design of the simulations and assumptions, it is possible to associate the value $\widetilde{r}$ to the components of $\vec{r}$, see Sec. 2.6.2.

For reconstructing the full discrete state, a probabilistic scheme was devised. It starts by constructing an undamaged lattice from which certain beams are removed, such that the resulting damage matches in a statistical sense the one given by $\vec{\kappa}$. Due to the assumed decoupling between the $x$- and $y$-zone, it is possible to handle the two directions independently. For each direction $\alpha$, i.e. $x$ and $y$, the following steps must be done:

1. From the continuum state $\kappa_\alpha$ the corresponding discrete state variable, $r_\alpha = \widehat{r}_\alpha(\kappa_\alpha)$ is computed. Through the relationship $N_\alpha := r_\alpha \cdot N_T$, it is possible to determine how many failed beams are associated to this direction.

2. Each beam is assigned a probability defined as
$$p_i \propto \frac{1}{\varepsilon_{i;th}} \left| \left\langle \vec{b}_i, \vec{t}_\alpha \right\rangle \right|^k, \qquad (9)$$
where $\varepsilon_{i;th}$ is the elongation threshold and $\vec{b}_i$ the direction of the beam. The vector $\vec{t}_\alpha$, called "damage basis", represents the main damage direction. In our motive, it is either $\vec{t}_x := (1, 0)^\mathrm{T}$ or $\vec{t}_y := (0, 1)^\mathrm{T}$. Finally, parameter $k$, called "directional weight", is a tuning parameter that balances the relative importance of the two terms and needs to be determined (see Sec. 3.4).

3. The $N_\alpha$ many beams to fail are drawn from the probability distribution defined by Eq. (9) without replacement.

4. The selected beams are marked as failed.

## 2.6. Numerical Simulations

For estimating and testing the damage function $\widehat{\mathbf{D}}(\vec{\kappa})$ and the transfer function $\vec{\widehat{r}}(\vec{\kappa})$, a series of different numerical simulations are carried out on fully discrete lattices. Due to the randomness of the lattice, 30 realisations were made for each case.

### 2.6.1. Numerical Simulation Procedure

The numerical simulations were carried out using a customised version of the Akantu FEM library (Richart et al., 2015). At the beginning of each loading step the corresponding boundary conditions are applied to the lattice. Then the following iterative procedure is carried out:

First, the equilibrium positions of the nodes is determined, without considering beam failure. Then, for each beam that has not yet failed the failure condition Eq. (4) is evaluated. Among the beams that satisfy the failure condition the one with the largest $\Psi$ value is removed. These steps are repeated until no beam satisfy the failure condition and the load increment ends.

### 2.6.2. The UniformSim Simulation Setup

The first type of simulation, called UniformSim, is used for estimating the transfer function $\vec{\widehat{r}}(\vec{\kappa})$ and the damage law $\widehat{\mathbf{D}}(\vec{\kappa})$. These simulations realise an uni-axial strain state of the lattice that is also rotated by an arbitrary but constant angle $\varphi$, called the pull direction. (see Fig. 4). Thus
$$\boldsymbol{\varepsilon}_\varphi := \mathbf{R}_\varphi^\mathrm{T} \begin{pmatrix} \widehat{\varepsilon}_1 & 0 \\ 0 & 0 \end{pmatrix} \mathbf{R}_\varphi, \qquad (10)$$
where $\mathbf{R}_\varphi$ is the Givens rotation matrix for angle $\varphi$, is applied to the lattice's boundary. In each loading step, $\widehat{\varepsilon}_1$ is increased by 0.0001 until 0.005 is reached. The limit is chosen to ensure that no localisation will occur and that damage maintains its diffuse character. The influence of the strain increment on the damage evolution is small (Appendix C).

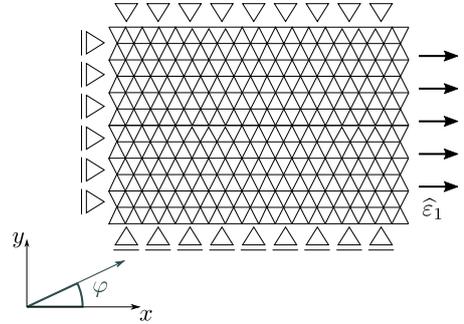

Figure 4: Boundary conditions used by the UniformSim series, shown for the case of $\varphi = 0°$. In general, the boundary conditions given by Eq. (10) are applied to the whole boundary. Scale of the lattice is exaggerated.

The particular setup of the UniformSim simulations together with the previous assumptions on $\mathbf{D}$ and $\vec{\widehat{r}}(\vec{\kappa})$ allows the following conclusions and simplifications:

(i) A pull direction is either associated to the $x$- or $y$-zone (see Sec. 2.2). This allows to probe the behaviour of a single zone. Which zone is probed depends on $\varphi$ and the yet unknown zone boundary value $\chi$.

(ii) A simulation, i.e. a particular value of $\varphi$, will only affect the state of either the $x$- or the $y$-zone. Thus, an increase of $\widehat{\varepsilon}_1$ will only affect one eigenvalue of $\mathbf{D}$ and a single component of $\vec{\kappa}$ as well as $\vec{r}$. Which component is affected depends on $\varphi$ and $\chi$.

(iii) For the continuum state variable the relation $\widetilde{\kappa} \stackrel{!}{=} \|\vec{\kappa}\|_1 \stackrel{!}{=} |\kappa_\alpha| \stackrel{!}{=} \widehat{\varepsilon}_1$ holds. Thus, one component equals the applied uni-axial strain $\widehat{\varepsilon}_1$, while the other is zero.

(iv) For the discrete state variable, the relation $\widetilde{r} \stackrel{!}{=} \|\vec{r}\|_1 \stackrel{!}{=} |r_\alpha|$ holds. Thus, only one component is non zero and equals $\widetilde{r}$. This can be used to determine the components of $\vec{r}$ from $\widetilde{r}$, once the $x$- and $y$-zones are known (see Sec. 3.4).

*2.6.3. The* `MultiLoadSim` *Simulation Setup*

For testing the damage function as well as the reconstruction procedure, a second type of simulation is used, called `MultiLoadSim`. It realises a bi-axial strain state, imposed along the $x$- and $y$-axes (see Fig. 5). Both strains are increased until $\varepsilon_{xx} = \varepsilon_{yy} = \varepsilon_{fin}$ is reached, where $\varepsilon_{fin}$ is the control parameter. For each simulation, the loading is imposed in three different ways, but each time the same initial lattice is used:

`XThenYSim`: $\varepsilon_{xx}$ is increased in steps of 0.0001 until it reaches $\varepsilon_{fin}$ and then maintained. Then $\varepsilon_{yy}$ is increased by the same increment until $\varepsilon_{fin}$ is reached.

`YThenXSim`: The same as `XThenYSim`, however, the order of loading the axes is switched.

`BothXYSim`: Both strains $\varepsilon_{xx}$ and $\varepsilon_{yy}$ are increased simultaneously, in steps of 0.0001 until $\varepsilon_{fin}$ is reached.

All three paths reach the same final state, $\varepsilon_{xx} = \varepsilon_{yy} = \kappa_x = \kappa_y = \varepsilon_{fin}$, but via different paths. As a consequence, the special relation $\widetilde{\kappa} := \|\vec{\kappa}\|_1 = |\varepsilon_{xx}| + |\varepsilon_{yy}|$ holds in these simulations. Both, the `XThenYSim` and the `YThenXSim` loading path impose in the first half of the loading an uni-axial strain state and then switch to a bi-axial strain state for the second half, while the `BothXYSim` loading path imposes a bi-axial strain state from the beginning.

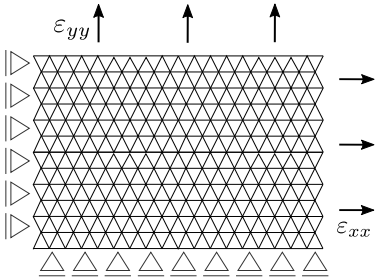

Figure 5: Boundary conditions used in the verification simulations. Scale of the lattice is exaggerated.

*2.6.4. The* `ReconstrSim` *Simulation Setup*

For testing the reconstruction process (see Sec. 2.5), a third type of simulation is used, called `ReconstrSim`. The basic setup is equivalent to `UniformSim`, but for $\varphi = 0°$. Further, a lattice with a certain initial damage is used. This damage was constructed to match the continuum state $\vec{\kappa} = (\widehat{\varepsilon}, 0)^{\mathrm{T}}$, *i.e.* the damage created by an uni-axial strain of $\widehat{\varepsilon}$ applied along the $x$-direction. Another important difference is, that the loading does not start at zero but at $\widehat{\varepsilon}$. In case of a perfectly working reconstruction process, one would expect no additional damage for strains $\leq \widehat{\varepsilon}$.

Unlike the `MultiLoadSim` tests, which focuses on the value of the reconstructed damage, these tests focus on how the reconstructed lattices behave after their reconstruction. In essence, this test simulates the exchange of the continuum representation with the discrete representation, *i.e.* the refinement process, which is the core application of the proposed method.

# 3. Results

Our proposed method relies on the damage law $\widehat{\mathbf{D}}(\vec{\kappa})$ used inside the continuum and the reconstruction process. First, we discuss how we will use the techniques introduced in Sec. 2 to process the data that we have collected from the numerical simulations. Then, we determine the damage law $\widehat{\mathbf{D}}(\vec{\kappa})$ and the zone boundary value $\chi$ (see Sec. 3.2) followed by an assessment of its accuracy (see Sec. 3.3). Thereafter, we repeat the process to determine the transfer function $\widetilde{\vec{r}}(\vec{\kappa})$ and the directional weight parameter $k$ (see Sec. 3.4). Finally, we demonstrate the applicability of our method (see Sec. 3.5).

*3.1. Details of a Numerical Simulation*

Let's consider a setting similar to `UniformSim` but with $\varphi = 0°$ (see Fig. 6a). Only one realisation is simulated and the loading goes beyond $\widehat{\varepsilon}_1 = 0.005$.

We measure the normalised stresses (solid lines) and compare them with the expected ones in case of suppressed damage, *i.e.* undamaged case (dashed lines) (Fig. 6b). As expected, initially the stresses behave predominantly linear. However, once a strain of about 0.003 is reached, we observe that $\widehat{\sigma}_{xx}$ starts to deviate from the undamaged case. While this deviation increases with further loading, we cannot observe it for $\widehat{\sigma}_{yy}$, that is much less affected by the loading. At one point, we observe that both stresses suddenly drop. This is caused by the emergence of a macroscopic crack, which is the expected behaviour for a brittle disordered material, such as concrete.

Using the techniques presented in Sec. 2.4, it is possible to extract the macroscopic damage variable $\mathbf{D}$ for the lattice, at any loading step. In Fig. 6c, we show the eigenvalues of the extracted damage variables where we can see that $d^{(x)}$ exceeds $d^{(y)}$. This also explains why $\widehat{\sigma}_{xx}$ deviates much more from the undamaged behaviour when compared to $\widehat{\sigma}_{yy}$. The underlying reason of this difference are the horizontal beams. They experience much larger strains than inclined beams, since they align with the loading and thus fail at much larger number. From Fig. 6c, we can also see that $d^{(y)}$ drops for a strain at around 0.002. This is a non-physical behaviour as damage should always increase. It is caused by some numerical issues during the determination of the stiffness tensor (see Sec. 2.4) and the sensitivity of damage extraction process.

Fig. 6d shows the ratio of failed beams (rfb), $\widetilde{r} := \|\vec{r}\|_1 := N_f/N_T$, where $N_f$ is the total number of number of failed beams and $N_T$ the total beams in the lattice. We will use it to indirectly estimate the transfer function $\widetilde{\vec{r}}(\vec{\kappa})$.

Figs. 6e-h show snapshots of the lattice's underlying microstructure. While taken at different loading steps (see Fig. 6d), they always show the same set of nodes, located

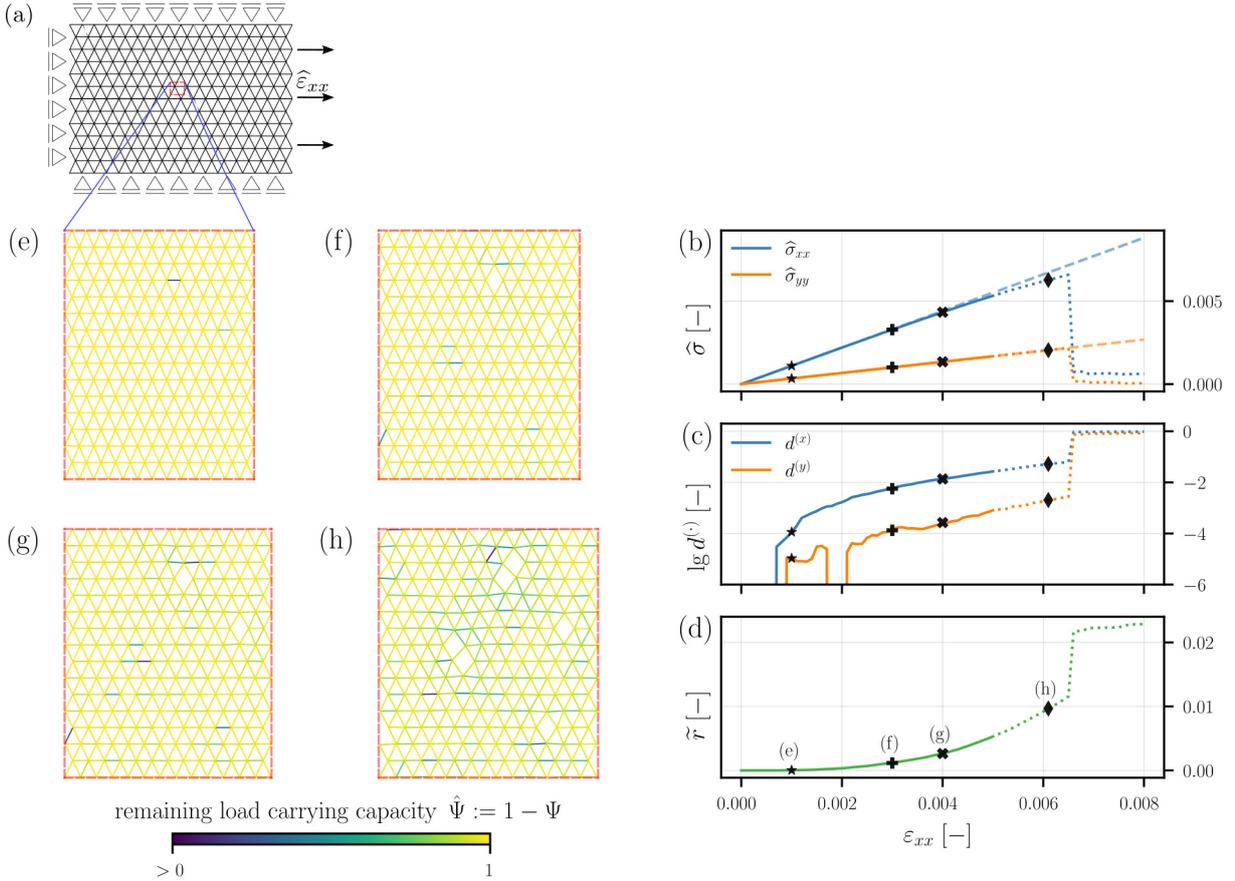

Figure 6: Representative simulation example. (a) Schematics of the model configuration, scale of the lattice is exaggerated. (b-d) Evolution of continuum and microstructure properties of the lattice. Dotted lines denote strains beyond the limit of 0.005 used in `UniformSim`. (b) Normalised measured stresses. Dashed lines represent the behaviour in case of suppressed damage. (c) Eigenvalues of the extracted damage variable **D**, see Eq. (1). (d) Ratio of failed beams $\widetilde{r}$ in the specimen. (e-h) Snapshots of a small section of the microstructure. Colours indicate the remaining load carrying capacity of the beams $\widehat{\Psi}_i := 1 - \Psi_i$, where $\Psi_i$ is defined by Eq. (4). Associated states are indicated in (d) by markers. Bending of beams is not shown.

roughly at the lattice's centre. While the exact damage pattern depends on the realisation of the lattice and locations where the snapshot was taken, statistically they all look the same. It is this statistical damage pattern that we want to capture by the transfer function $\overrightarrow{\widetilde{r}}(\vec{\kappa})$ and recreate by the reconstruction process. Whereas the damage law $\widehat{\mathbf{D}}(\vec{\kappa})$ captures the accumulated effects on the lattice's macroscopical stiffness.

### 3.2. Estimation of the Damage Law $\widehat{\mathbf{D}}(\vec{\kappa})$

We now study the behaviour of the damage variable **D** that we have extracted from the data of the `UniformSim` simulations. From these observations, we will determine the damage function $\widehat{\mathbf{D}}(\vec{\kappa})$ as well as the zone boundary value $\chi$ (see Fig. 2).

*Functional Form of $\widehat{d}_x(\kappa_x)$ and $\widehat{d}_y(\kappa_y)$.* Since we have assumed an orthotropic damage variable (see Sec. 2.2), we have to assume the same for the damage function. Thus arriving the tentative form of the damage function is given by

$$\widetilde{\mathbf{D}}(\vec{\kappa}) := \begin{pmatrix} d_{xx}(\vec{\kappa}) & 0 \\ 0 & d_{yy}(\vec{\kappa}) \end{pmatrix}.$$

To account for deviations from this assumption, we will connect the two diagonal elements of the damage function with the eigenvalues of the measured damage variable. Thus, we have only two functions that we need to determine.

Fig. 7 shows the evolution of the eigenvalues $d^{(x)}$ and $d^{(x)}$ from the extracted damage variable for the pull directions $\varphi \in \{0°, 60°\}$ at various distortion levels. For $\varphi = 0°$, the eigenvalue $d^{(x)}$ is much larger than $d^{(y)}$, while for $\varphi = 60°$ the opposite is observed. Later, we will use this to determine the zone boundary value $\chi$. Most importantly, the figures show that both eigenvalues follow a power law, irrespective of the pull direction and distortion. Thus we approximate the diagonal elements/eigenvalues of $\widehat{\mathbf{D}}(\vec{\kappa})$ as:

$$d_{xx} \approx \widehat{d}_x(\kappa_x;\, a,\, \varphi) := \alpha_{a,\varphi}^{(x)} \cdot \kappa_x^{\beta_{a,\varphi}^{(x)}}, \tag{11a}$$

$$d_{yy} \approx \widehat{d}_y(\kappa_y;\, a,\, \varphi) := \alpha_{a,\varphi}^{(y)} \cdot \kappa_y^{\beta_{a,\varphi}^{(y)}}. \tag{11b}$$

The parameters of these approximations depend on the distortion level $a$ and the pull direction $\varphi$. Later, we will eliminate their dependency on $\varphi$ and obtain the final parameters that only depend on $a$, which is constant. Fur-ther, this choice guarantees that the damage is strictly increasing.

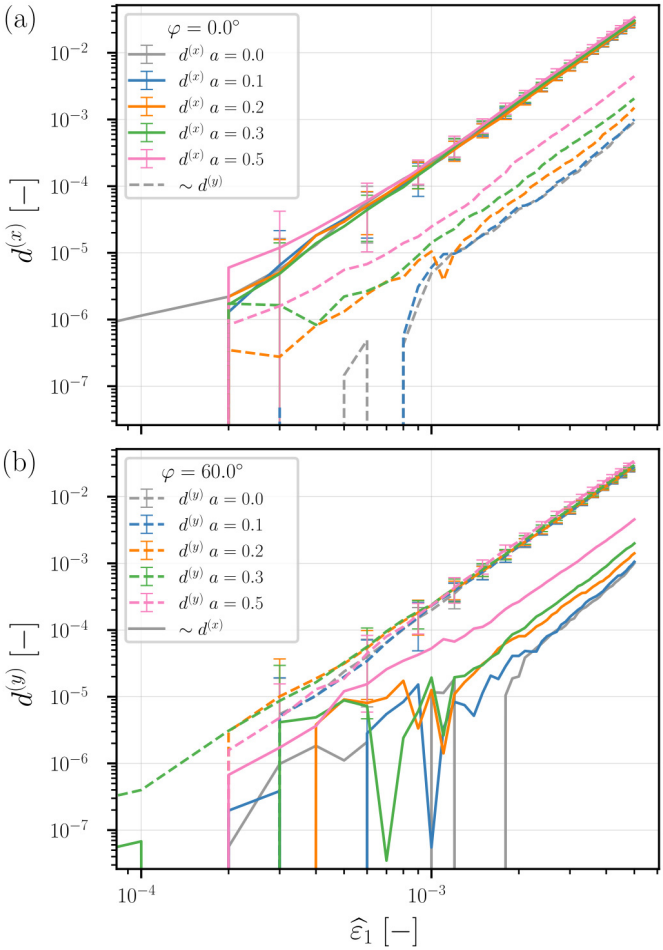

Figure 7: Eigenvalues of the extracted damage variable $\mathbf{D}$ for pull directions $\varphi = 0°$ (a) and $60°$ (b). Solid lines correspond to $d^{(x)}$, while dashed lines correspond to $d^{(y)}$. Colours indicate different distortion levels $a$ of the underlying lattice.

Because of our previous assumption about the independence of the directions, the approximations of the eigenvalues only depend on a single component of the continuum state $\vec{\kappa}$ (Sec. 2.2). While this could be justified due to their large differences, that we can see in Fig. 7, we clearly see that even for $\varphi = 0°$, there is a certain coupling between $d^{(x)}$ and $d^{(y)}$. To handle this, we use a simple coupling scheme, which leads to the final damage function

$$\widehat{\mathbf{D}}(\vec{\kappa}) := \qquad (12)$$
$$\begin{pmatrix} \max\left\{\widehat{d}_x(\kappa_x), \frac{\widehat{d}_y(\kappa_y)}{\eta}\right\} & 0 \\ 0 & \max\left\{\widehat{d}_y(\kappa_y), \frac{\widehat{d}_x(\kappa_x)}{\eta}\right\} \end{pmatrix},$$

where $\widehat{d}_x(\kappa_x)$ and $\widehat{d}_y(\kappa_y)$ are the approximations of the eigenvalues defined by Eq. (11) but without the dependence on $\varphi$. The coupling ensures that the eigenvalues of the damage function $\widehat{\mathbf{D}}(\vec{\kappa})$ will at most differ by a factor of $\eta$, which is exactly what we see in the case of uni-axial loading (see Fig. 7). Here, we will assume that the empirical parameter $\eta$ equals 10 in all cases. We will later give a justification of the form and value of the proposed coupling. It is important to notice that this coupling is designed for the uni-axial case. However, a more elaborated coupling might be needed, depending on the details of other material motives.

*Parameters of $\widehat{d}_x(\kappa_x)$ and $\widehat{d}_y(\kappa_y)$.* Since the data, especially for the non-dominant eigenvalue shows strong variation for small strains, only data points corresponding to strains above 0.002 were used for the parameter estimation. In Figs. 8a,b, we see that for small values of $\varphi$, the $\alpha_{a,\varphi}^{(x)}$-parameters are very close to each other, while for larger values of $\varphi$ one observes a much larger scattering.

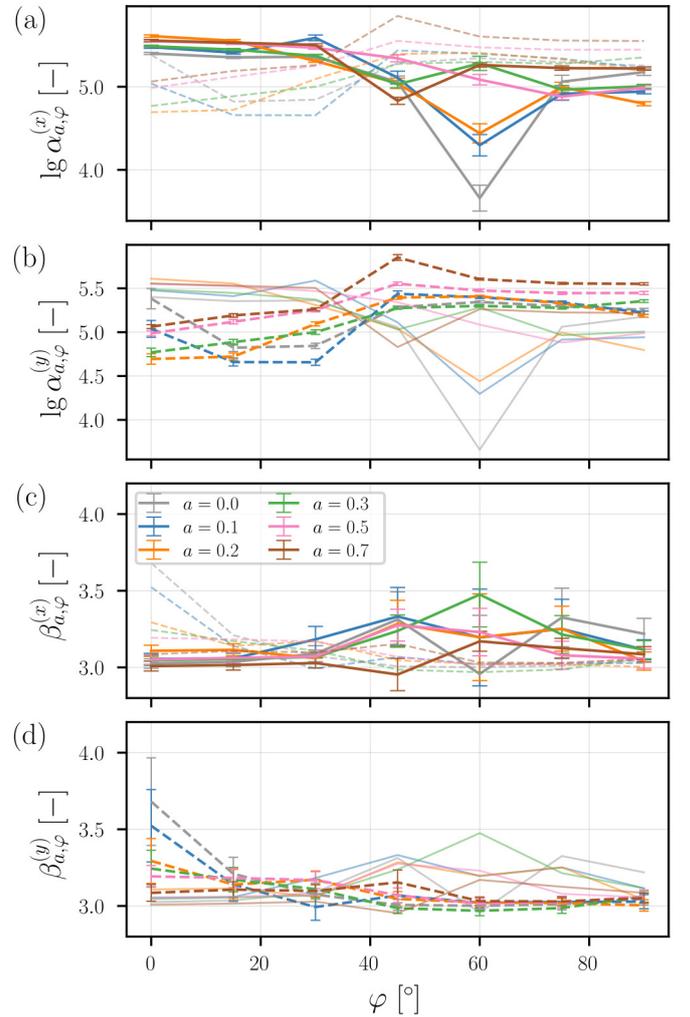

Figure 8: Values of the $\alpha$- (a,b) and $\beta$-parameters (c,d) with respect to the pull direction $\varphi$. Colours indicate different distortion levels. Solid lines correspond to $\lg \alpha_{a,\varphi}^{(x)}$ and $\beta_{a,\varphi}^{(x)}$, while dashed lines to $\lg \alpha_{a,\varphi}^{(y)}$ and $\beta_{a,\varphi}^{(y)}$. Error bars indicate the 95% confidence interval.

Interestingly, $\alpha_{a,\varphi}^{(y)}$-parameters behave inversely. Furthermore, on Fig. 8b we can clearly observe the $\alpha_{a,\varphi}^{(y)}$ dependence on $\varphi$. We see that $\alpha_{a,\varphi}^{(y)}$ is small if $\varphi$ is small too, but above a certain value of $\varphi$, the parameters become much larger and their scattering increases. The same, but in an opposite way, holds for the $\alpha_{a,\varphi}^{(x)}$-parameters but in a less pronounced fashion.

The estimates for the $\beta$-parameters (see Figs. 8c,d) show a similar behaviour with respect to $\varphi$. However, while we observed a significant change in the behaviour of the $\alpha$-parameters' values, from a particular value of $\varphi$ on we just observe an increase of the variability of $\beta$.
In summary, from Fig. 8 we can conclude that the $\beta$- and especially the $\alpha^{(y)}$-parameters have different regimes depending on $\varphi$. Further, inside such a regime, their particular value does not depend much on $\varphi$.

We also saw that the values for the $\beta^{(x)}$-parameters for small values of $\varphi$ and $\beta^{(y)}$-parameters for large values of $\varphi$ are both close to three. This means that the growth behaviour of $\widehat{d}_x(\kappa_x)$ and $\widehat{d}_y(\kappa_y)$ are very similar. This justifies the form of the coupling used in the damage function in Eq. (12).

*Zone Boundary $\chi$.* In Figs. 7 and 8, we have observed that depending on the pull direction either $d^{(x)}$ or $d^{(y)}$ is dominant. We now exploit this fact to define $\chi$. To this end, we define the dominance function $\zeta$ as:

$$\zeta(a, \varphi) := \lg\left(\frac{d_{a,\varphi;\,\tilde{\kappa}=0.005}^{(x)}}{d_{a,\varphi;\,\tilde{\kappa}=0.005}^{(y)}}\right), \quad (13)$$

with $d_{a,\varphi;\,\tilde{\kappa}=0.005}^{(\alpha)}$ as the damage eigenvalue associated to direction $\alpha$, once the uni-axial strain has reached 0.005. The most important aspects of this function are its sign and root, to a lesser extend its value. $\zeta > 0$ means that $d^{(x)}$ is dominant, while $\zeta < 0$ indicates that $d^{(y)}$ is dominant. Thus, $\chi$, which might depend on the distortion $a$, is defined as $\zeta(a, \chi) \stackrel{!}{=} 0$.
Examining Fig. 9, we see that, irrespective of the distortion, $\chi$ must lie between 30° and 45°. After some experimentation, we decided to use 40° as zone boundary, irrespective of the distortion level. A closer analysis might yield different estimations.

$\zeta$ can be seen as a measure of the coupling between $d^{(x)}$ and $d^{(y)}$. Thus, we can used it to determine the value of the empirical coupling parameter $\eta$, see Eq. (12). Our value $\eta = 10$ was selected because it is roughly the mean value for $\varphi = 0°$.

*Final Parameters of $\widehat{d}_x(\kappa_x)$ and $\widehat{d}_y(\kappa_y)$.* Eliminating the dependency of the $\alpha$- and $\beta$-parameters on the pull direction $\varphi$ will results in parameters that are valid inside the entire $x$- or $y$-zone. For this, we combine the different

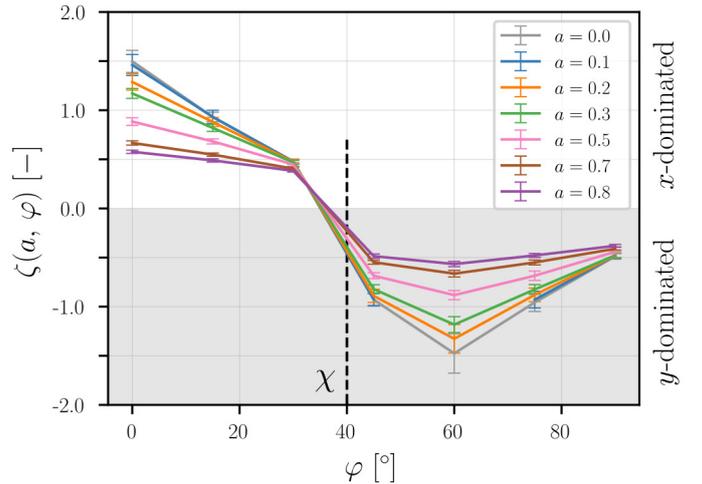

Figure 9: Dominance function $\zeta(a, \varphi)$, Eq. 13, for different distortion parameters $a$. The $x$-dominated region, *i.e.* $d^{(x)} \gg d^{(y)}$, is defined by $\zeta > 0$, while the $y$-dominated (grey shaded) region, *i.e.* $d^{(x)} \ll d^{(y)}$, is defined by $\zeta < 0$. The `UniformSim` data was used.

estimates as:

$$\lg \alpha_a^{(x)} := \frac{1}{|\mathcal{X}|}\sum_{\varphi \in \mathcal{X}} \lg \alpha_{a,\varphi}^{(x)}, \quad \lg \alpha_a^{(y)} := \frac{1}{|\mathcal{Y}|}\sum_{\varphi \in \mathcal{Y}} \lg \alpha_{a,\varphi}^{(y)}, \quad (14a)$$

$$\beta_a^{(x)} := \frac{1}{|\mathcal{X}|}\sum_{\varphi \in \mathcal{X}} \beta_{a,\varphi}^{(x)}, \qquad \beta_a^{(y)} := \frac{1}{|\mathcal{Y}|}\sum_{\varphi \in \mathcal{Y}} \beta_{a,\varphi}^{(y)}, \quad (14b)$$

where $\mathcal{X}$ contains all the pull directions associated to the $x$- and $\mathcal{Y}$ the ones associated to the $y$-zone. Parameters associated to the transversal directions are simply ignored, *e.g.* $\lg \alpha_{a,\varphi=0°}^{(y)}$. Further, the functional form of $\widehat{d}_x(\kappa_x)$ and $\widehat{d}_y(\kappa_y)$ is still given by Eq. (12), just without the dependency on $\varphi$. Note that Eq. (14) weights the different pull directions equally.

### 3.3. Test of the Damage Evolution Law $\widehat{\mathbf{D}}(\vec{\kappa})$

We now evaluate how well the damage function is able to predict the damage of a fully discrete simulation. For this purpose, the `MultiLoadSim` simulations are used.

Fig. 10 shows the results of such an experiment for $\varepsilon_{fin} = 0.002$. We can see the eigenvalues, once computed for the reference (solid), *i.e.* a fully discrete simulation, and alternatively computed by the damage function $\widehat{\mathbf{D}}(\vec{\kappa})$ (dash-dotted), *i.e.* CDM. They are plotted against the normalised total strain $\tau := \tilde{\kappa}/2\varepsilon_{fin}$. Thus, both the `X-ThenYSim` (orange) and the `YThenXSim` (green) load paths switch from an uni-axial to a bi-axial strain state at $\tau = 0.5$. We observe that irrespective of the loading path the same final damage values are reached. The value depends on the used method, since the final value of the CDM is different from the reference value. Note that this is not problematic since the CDM is only used during the initial phase.

Fig. 10 also demonstrates that the damage for the `X-ThenYSim` and `YThenXSim` are very similar to each other.

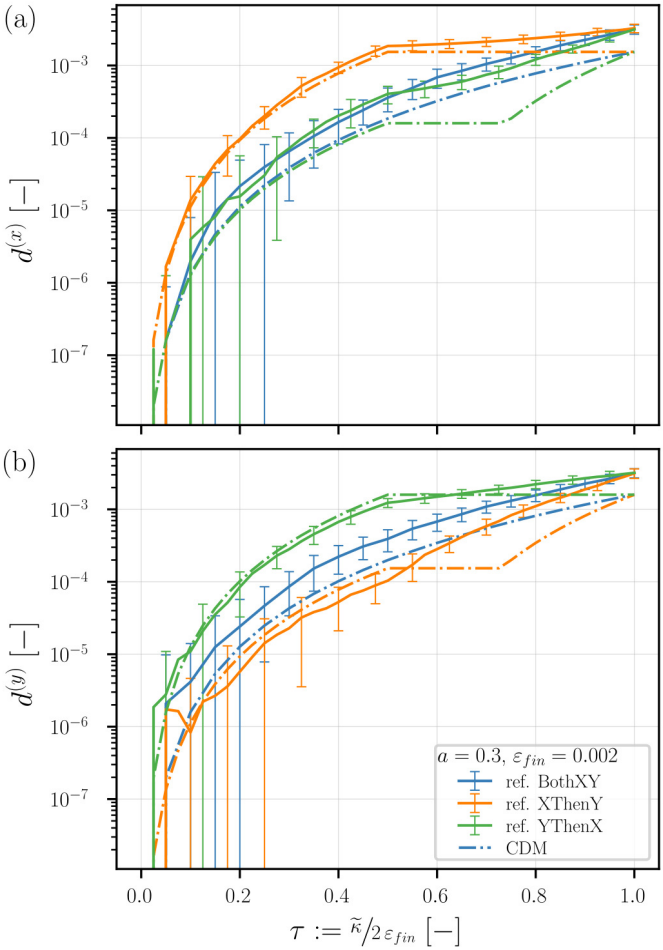

Figure 10: Damage eigenvalues for the three different loading paths, described in Sec. 2.6.3, with final strain $\varepsilon_{fin} = 0.002$, plotted against $\tau := \tilde{\kappa}/2\,\varepsilon_{fin}$. Using a fully discrete simulation (solid) as reference and the CDM damage law $\widehat{\mathbf{D}}(\vec{\kappa})$ (dash-dotted). The colours indicates the three different loading paths. The distortion of the lattices was $a = 0.3$.

However, the eigenvalues are flipped, which is the expected behaviour. During the first half of the loading (i.e. $\tau < 0.5$), the prediction of the dominant eigenvalue matches well with the reference value for both loading paths. At the same time, the non-dominant eigenvalue, i.e. the one belonging to the transverse direction, is captured with less but still acceptable accuracy. The mismatch is entirely due to the rather crude choice of the $\eta$ coupling parameter (see Eq. (12)). However, it indicates that the proposed coupling is indeed working.

Nevertheless, for the second half of the loading (i.e. $\tau > 0.5$) the CDM is unable to capture the evolution to a satisfactory degree. In case of `XThenYSim` (orange lines), we see that the CDM approximation of the $x$-eigenvalue $\widehat{d}_x$ remains constant, since $\kappa_x$ is not affected by a loading along the $y$-axis. However, we see that in the reference system $d^{(x)}$ continuously increase (see Fig. 11 for more). The $y$-eigenvalue $\widehat{d}_y$ predicted by the CDM remains initially con-stant due to the coupling. Once $\widehat{d}_y(\kappa_y)$ has become larger than $\widehat{d}_{x(\varepsilon_{fin})}/\eta$ $\widehat{d}_y(\kappa_y)$ starts to increase. However, as it can be seen form Fig. 10b, the reference $d^{(y)}$ starts to increase almost immediately.

A different case is the `BothXYSim` loading path. From Fig. 10, it seems that for $\tau < 0.5$ its damage grows slower than the dominant damage observed for the other two paths. This is because `BothXYSim` only has half the numbers of loading steps the other two have. If this is corrected for then it would actually grow faster. This indicates that there is some form of coupling between the two directions that is not considered correctly.

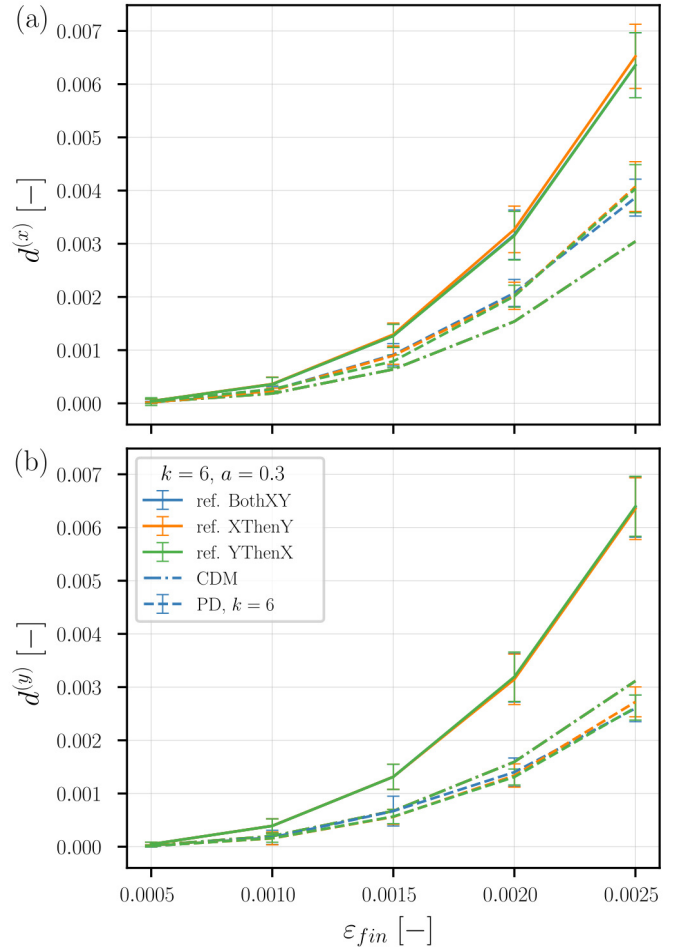

Figure 11: Final value of the $d^{(x)}$ and $d^{(y)}$ damage eigenvalues, computed using the reference (solid), CDM (dash doted) and reconstruction (dashed) method, plotted against $\varepsilon_{fin}$. The colours indicate the three different loading cases from Sec. 2.6.3. All lattices have a distortion of $a = 0.3$.

In Fig. 11, we can see how the final damage, i.e. values of $d^{(x)}$ and $d^{(y)}$ at $\varepsilon_{xx} = \varepsilon_{yy} = \varepsilon_{fin}$, depend on the control parameter $\varepsilon_{fin}$, using either the reference (solid lines), the CDM (dash-dotted lines) or the reconstruction (dashed lines). The colours distinguish the three different load paths that were tested (see Sec. 2.6.3). The collapse of the lines indicate that the damage is indeed path in-dependent, regardless of the final strain $\varepsilon_{fin}$. However, the final value depends on the particular method that was used. In in Fig. 10, we observe a gap between the final damage attained by the reference and the one predicted by the CDM. We can now see that this gap is systematic and actually increases with larger $\varepsilon_{fin}$. This is again indicating that there is some form of coupling between the directions that is not take into account yet.

### 3.4. Estimation of the Transfer Function $\vec{\hat{r}}(\vec{\kappa})$

Our procedure to reconstruct a discrete lattice representation based on a damaged continuum state (presented in Sec. 2.5) requires two unknown quantities: First, the transfer function $\vec{\hat{r}}(\vec{\kappa})$, which maps the continuum state $\vec{\kappa}$ to the corresponding discrete surrogate state $\vec{r}$. Second, the directional weight parameter $k$, which balances the orientation and the strength of a beam during the reconstruction process (see Eq. (9)). Analogously to the determination of the damage function, the data from the `UniformSim` is used.

*Functional Form of $\hat{r}_x(\kappa_x)$ and $\hat{r}_y(\kappa_y)$.* As mentioned before, it is impossible to measure the components of $\vec{r}$ directly. However, as outlined in Sec. 2.6.2 $\vec{r}$ is connected to the ratio of failed beam as $\widetilde{r} = \|\vec{r}\|_1 = {N_f}/{N_T} \stackrel{!}{=} r_\alpha$. Thus, we can estimate $\vec{r}$ indirectly. Fig. 12 shows $\widetilde{r}$ for the pull direction $\varphi = 0°$ at various distortion levels. As we can see, the distortion level has only minor influence. Different pull directions do not lead to a qualitative change (data not shown). For that reason, we approximate the mean rfb as

$$\widetilde{r} \approx \left\|\vec{\hat{r}}(\vec{\kappa})\right\|_1 := \widehat{r}(\kappa;\, a,\, \varphi) := \alpha^{(r)}_{a,\varphi} \cdot \kappa^{\beta^{(r)}_{a,\varphi}}, \quad (15)$$

with the two fit parameters $\alpha^{(r)}_{a,\varphi}$ and $\beta^{(r)}_{a,\varphi}$. Both depend on the distortion $a$ and the pull direction $\varphi$. Due to our previous assumptions, we can identify its argument $\kappa$ directly with $\widetilde{\kappa}$. To eliminate the dependency on $\varphi$ we use the same method as for the damage function (see Sec. 3.2). However, parameters associated to pull directions in $\mathcal{X}$ are used to determine $\widehat{r}_x(\kappa_x)$, while the ones belonging to $\mathcal{Y}$ determine $\widehat{r}_y(\kappa_y)$. This will transform the approximation of the scalar quantity $\left\|\vec{\hat{r}}(\vec{\kappa})\right\|_1$ into the one for $\vec{\hat{r}}(\vec{\kappa})$.

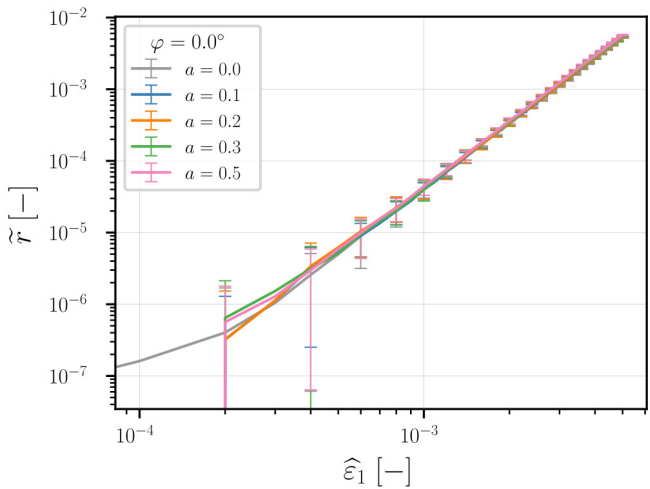

Figure 12: $\langle \tilde{r} \rangle := \langle \|\vec{r}\|_1 \rangle$ for pull direction $\varphi = 0°$, at different values of the distortion parameter $a$. No qualitative change is observed for different pull directions $\varphi$.

For the discussion about the estimated $\alpha$- and $\beta$-parameters please see Appendix A.

*Directional Weight Parameter $k$.* The empirical tuning parameter $k$ influences the selection of beams during the reconstruction process. It balances a beam's strength, *i.e.* its elongation threshold $\varepsilon_{th}$, against how well it aligns with the damage basis $\vec{t}_\alpha$ (see Sec. 2.5). We determine $k$ such that the reconstructed damage variable $\overline{\mathbf{D}}$ resembles the reference damage $\mathbf{D}$ most closely. To this end, we define

$$\Upsilon_k := \left\|D_{11} - \overline{D}_{11}\right\|_{\ell_2} + \left\|D_{22} - \overline{D}_{22}\right\|_{\ell_2} + 2\left\|D_{12} - \overline{D}_{12}\right\|_{\ell_2} \quad (16)$$

as a measure of separation between the two damages. For minimising $\Upsilon_k$, we select a heuristic approach, in which the reconstruction process (see Sec. 2.5) is run for different values of $k$. The $k$ that minimises $\Upsilon_k$ will then be used for the remaining part of this paper. However, for this particular reconstruction process, the used $\alpha^{(r)}$- and $\beta^{(r)}$-parameters still depended on $\varphi$. Further, zoning was ignored and as damage basis the pull direction $\varphi$ was used. The underlying lattice was not distorted. From Fig. 13 we see that $k = 6$ minimises $\Upsilon_k$ independent of the pull direction. We also see that pull directions $\{30°, 90°\}$ seem to be almost unaffected by $k$, however, their values match $\Upsilon_{k=6}$. This is an artefact caused by the regular structure of the underlying lattice and the scalar product used in the definition of the selection probability (see Eq. (9)). However, this artefact is indicating that $k = 6$ is indeed a good choice.

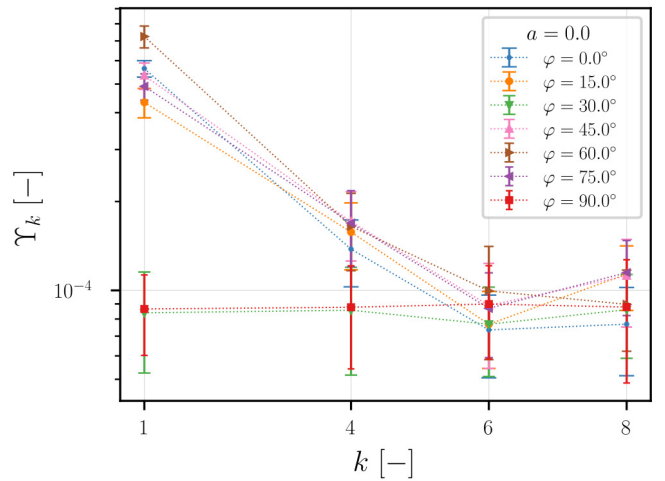

Figure 13: $\Upsilon_k$, Eq. (16), for some values of $k$ and various pull directions. The reconstruction process was done on regular grids, without zoning. Further the pull direction and its orthogonal was used as damage basis.

## 3.5. Tests of the Reconstruction Process

Now we evaluate the performance of the proposed reconstruction scheme to create a mechanically equivalent lattice, based solely on the continuum state $\vec{\kappa}$ (see Sec. 2.5). For verification, we use the `MultiLoadSim` simulations (see Sec. 2.6.3). In addition, we use the `ReconstrSim` simulations to simulate a refinement step (see Sec. 2.6.4).

*The `MultiLoadSim` Results.* In Fig. 14, we see the results from the `MultiLoadSim` setup with $\varepsilon_{fin} = 0.002$ (see Sec. 2.6.3). They impose the bi-axial strain state $\varepsilon_{xx} = \varepsilon_{yy} = \varepsilon_{fin}$, but with loading applied via three different paths. We used this setup before to assess the CDM (see Sec. 3.3). An important note concerning the reconstructed states is, that in each loading step the lattice and hence the damage is constructed anew. Thus, although it looks like a damage evolution, the damage at any loading step has no connection to the previous one. However, each time the same undamaged but distorted lattice was used.

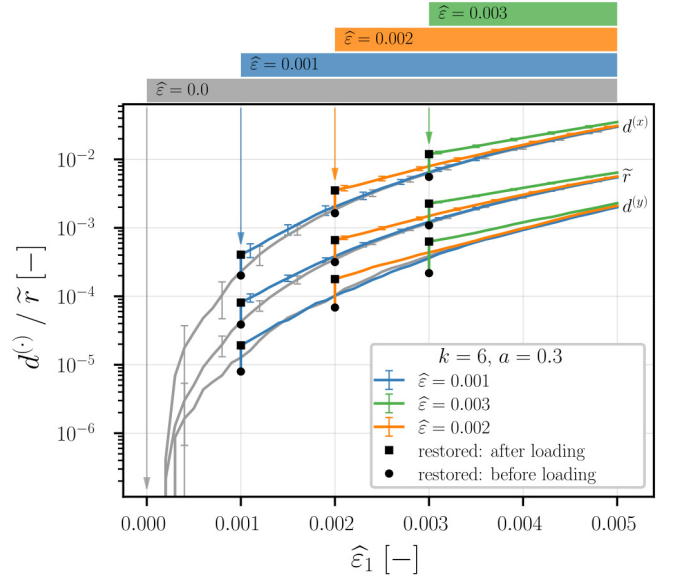

Figure 15: Behaviour of the $d^{(x)}$ and $d^{(y)}$ damage eigenvalues and $\widetilde{r}$. Colours indicate different reconstruction parameters $\widehat{\varepsilon}$. Grey is the reference, i.e. no reconstruction. Circles indicate damage/rfb values of the lattices directly after reconstruction. Squares indicate damage/rfb values of the lattices for an applied strain of $\widehat{\varepsilon}$.

If we now compare the damage from the references (solid lines) with the one from the reconstructed lattices (dashed lines) in Fig. 14, we see that the overall damage values are very similar to the ones obtained by the CDM. As before, we observe that for $\tau < 0.5$ the dominant eigenvalues, i.e. $d^{(x)}$ for `XThenYSim` and $d^{(y)}$ for `YThenXSim`, are captured well. Then, for $\tau > 0.5$, these reconstructed eigenvalues stop growing and thus deviate from the references (solid lines). An effect we observed for CDM (dashed lines), too. But if we look at the other eigenvalues, i.e. $d^{(y)}$ for `XThenYSim` (orange) and $d^{(x)}$ for the `YThenXSim` (green), we see that they start to increase almost immediately like the reference. This was not the case for the CDM (dash-dotted lines shown in Fig. 10). The reconstruction process is affected by the ignored coupling between the directions as well. However, the damage eigenvalues generated by it follow the reference much better than the ones computed by the CDM.

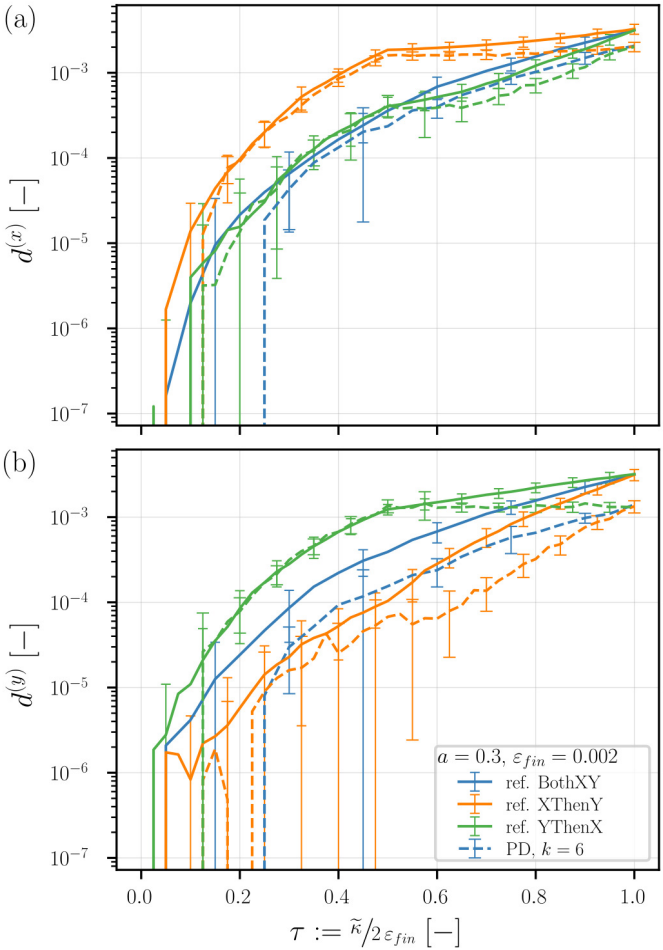

Figure 14: Damage eigenvalues for the three different loading paths, described in Sec. 2.6.3, with final strain $\varepsilon_{fin} = 0.002$, plotted against $\tau := \tilde{\kappa}/2\varepsilon_{fin}$. Using fully discrete simulations (solid) as reference and the reconstructed damage (dashed). The colours indicate the three different loading paths that were taken. The distortion of the lattices was $a = 0.3$. See Fig. 10 for the damage evolution predicted by the CDM.

*The `ReconstrSim` Results.* Now we are using the `ReconstrSim` simulation setup, described in Sec. 2.6.4. The lattices that are used here were reconstructed for the continuum state $\vec{\kappa} = (\widehat{\varepsilon}, 0)^{\mathrm{T}}$. The system is loaded under uni-axial strain along the $x$-axis, starting at $\widehat{\varepsilon}$. This setup simulates how a discrete region that was loaded up to $\widehat{\varepsilon}$, as continuum and then refined behaves upon further loading.

In Fig. 15, we see how the damage eigenvalues $d^{(x)}$ and $d^{(y)}$ (dashed and dotted lines, respectively) and the rfb $\widetilde{r}$ (solid lines), evolve for different reconstruction strains $\widehat{\varepsilon}$, indicated by different colours. The grey lines correspond to the reference without any reconstruction.

The circles in Fig. 15 indicate the values for $d^{(x)}$, $d^{(y)}$ and $\widetilde{r}$ have in the reconstructed lattice before any loading was applied to them. The circles associated to $\|\vec{r}\|_1$ show, that the reconstructed lattices have a matching rfb value $\widetilde{r}$ which is a consequence of its construction. It is, however, much more important, that the reconstructed $d^{(x)}$ eigenvalue (circles), matches the one predicted by the reference. Thus the process is able to reconstruct the dominant eigenvalue.

We are also observing that $d^{(y)}$ are not as well reconstructed. This is a consequence of the assumptions that the components of $\vec{r}$ are independent. Since `ReconstrSim` only impose strains along the $x$-axis, we have $\kappa_y \equiv 0 \Rightarrow r_y \equiv 0$. Thus, the reconstructed $y$-eigenvalues we are seeing are caused by a directional sampling effect created during the reconstruction of the damage. However, since $d^{(y)}$ is the non-dominant eigenvalue, we expect and allow that it is less well reconstructed.

The squares in Fig. 15 indicate the state of the lattices after an uni-axial strain of $\widehat{\varepsilon}$ along the $x$-axis was applied to them. The difference between a square and its associated circle proves that this strain causes the failure of additional beams. If the reconstruction process would work perfectly, any strain below or equal $\widehat{\varepsilon}$ should not lead to the failure of any beam. Therefore, we might have removed the right number of beams and these were more or less correctly oriented, the selection of some of them was not fully optimal.

Furthermore, we see that for subsequent loading steps, the damage and rfb continue to increase (Fig. 15). While the observed values for the restored systems remain above the reference, the restored lattices slowly converge towards them. This is because the damage created by the subsequent loading steps starts to dominate the artificial one, that we created through the restoration process.

## 4. Summary and Conclusion

In this study, we presented a generic approach for the creation of a discrete twin of a continuum representation containing an initial damage. The discrete twin's damage is created in such a way, that it is mechanically consistent to the original's continuum damage. This is a step towards adaptive multi-scale simulations, which take the state of the coarse description of a region into account upon its refinement.

While the method is general and has no restrictions concerning the used numerical representations, we presented it in form of a concrete example. As continuum representation, we have used FEM with CDM as damage measure. For the discrete representation, we have used a lattice based on a triangular grid consisting of brittle beam-truss elements.

One part of our method is the damage measure used inside the continuum representation. This measure is used during the initial continuum phase to track the evolving continuum damage. Unlike classical CDM, that are calibrated to match the degradation of a particular material, we calibrated the CDM against the degeneration of the discrete numerical representation. Thus, it measures the degeneration that would occur on a hypothetical fine scale representation.

We saw that the determined CDM is indeed able to capture the damage caused by uni-axial strains to a satisfying degree. However, for bi-axial loading, the CDM is unable to achieve the same. This is explained by the assumption that the directions are independent. Directional coupling must be used to further improve the CDM's accuracy

The second part of our method is the ability to construct a discrete damage that is mechanically consistent to a given continuum state $\vec{\kappa}$. Since this problem is obviously not unique, we devised a stochastic scheme to generate representations containing such a particular discrete damage.

We have seen, that the reconstruction process is indeed able to create discrete lattices, whose initial degeneration is consistent with the given continuum state $\vec{\kappa}$. A drawback is, that imposing a strain that corresponds to $\vec{\kappa}$ itself, leads to the failing of some additional beams. This is indicating that the selection process needs to be further refined. Furthermore, as for the CDM, we observed problems for bi-axial strains, which are again caused by the independence assumed between the directions.

Nevertheless, our data is indicating that our approach works well for the case of uni-axial loading and is in principal able to work for bi-axial loading. The next step is to integrate our method into an adaptive multi-scale simulation scheme.

## 5. CRediT

Philip Müller: Conceptualisation, Methodology, Software, Validation, Writing - Original Draft, Visualisation; Falk Wittel: Conceptualisation, Writing - Review & Editing, Supervision; David Kammer: Conceptualisation, Writing - Review & Editing, Supervision.

## 6. Declaration of Competing Interest

The authors declare that they have no known competing financial interests or personal relationships that could have appeared to influence the work reported in this paper.

## 7. Data Availability

The simulation data generated in this study have been deposited in the ETH Research collection database available at `TBA`.

# Appendix A. Parameters of $\widehat{r}_x(\kappa_x)$ and $\widehat{r}_y(\kappa_y)$

The fitting parameters (see Eq. (15)) of the ratio of failed beams (rfb) $\widetilde{r}$, denoted as $\alpha^{(r)}_{a,\varphi}$ and $\beta^{(r)}_{a,\varphi}$ were estimated in the same way as the ones for the two damage functions $\widehat{d}_x(\kappa_x)$ and $\widehat{d}_y(\kappa_y)$. Figs. A.16a show the values for the $\alpha$- and A.16b for the $\beta$-parameters. Compared with the parameters we obtained for the damage law (see Fig. 8) we see much less variability here. This is because it is far easier to measure this quantity than the damage.

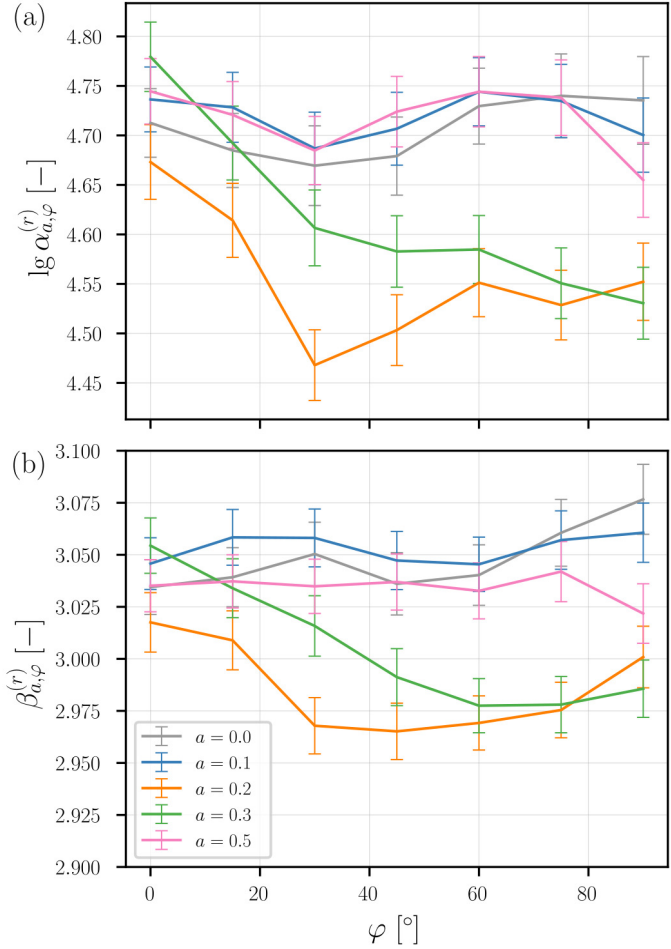

Figure A.16: Dependence of the two fitting parameters of Eq. (15), $\lg \alpha^{(r)}_{a,\varphi}$ (a) and $\beta^{(r)}_{a,\varphi}$ (b), on the pull direction $\varphi$. Colours indicating different distortion levels $a$. The error bars is given by the 95% confidence interval.

# Appendix B. Influence of the Characteristic Length $\ell$

The characteristic size is defined as $\ell := L_x/(N_x - 1)$, where $L_x$ is the specimen's $x$-extension and $N_x$ the number of nodes along the $x$-direction. We will now study its influence on the observed damage. For this a simulation similar to `UniformSim` with $\varphi = 0°$ was performed for different values of $N_x$. To account for the randomness the results are averaged over 30 runs. In Fig. B.17 we observe that, with the exception of the very coarse $N_x = 50$ grid, all lines collapse. This shows that the damage evolution is independent from the discretisation.

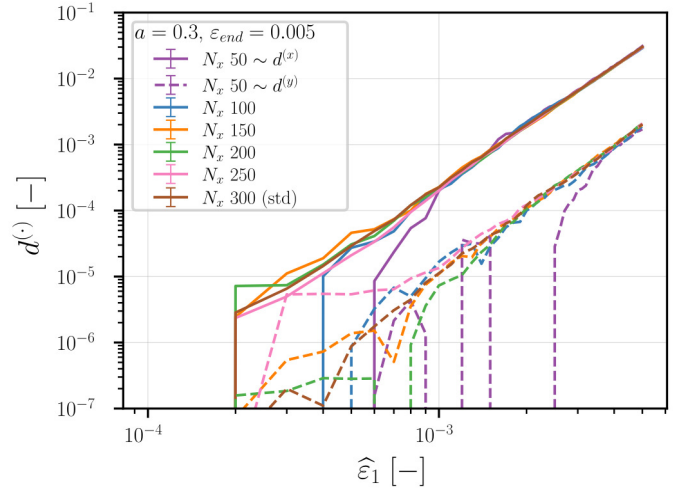

Figure B.17: Evolution of the damage eigenvalues $d^{(x)}$ (solid line) and $d^{(y)}$ (dashed line) for distortion level $a = 0.3$. The specimen was strained along the $x$-direction up to $\varepsilon_{end} = 0.005$. Different colours indicate different $N_x$, number of nodes along the $x$-direction. $N_x = 300$ is the default case, see Table I. For all cases the standard deviation (omitted) is roughly the same, see Fig. 7.

# Appendix C. Influence of the Number of Loading Steps

In the simulation described in Sec. 2.6, the loading is increased in steps of $10^{-4}$, until the final strain of $\varepsilon_{end} = 0.005$ is reached, which results in 50 increments. The effect of the loading step is studied with a series of simulations similar to `UniformSim` with $\varphi = 0°$. The specimen is loaded until a strain of $\varepsilon_{end} = 0.005$ is reached, but with a different number of load increments. To account for the variation each simulation was performed 30 times. In Fig. C.18 we observe that the damage evolution is in-dependent of the loading rate.

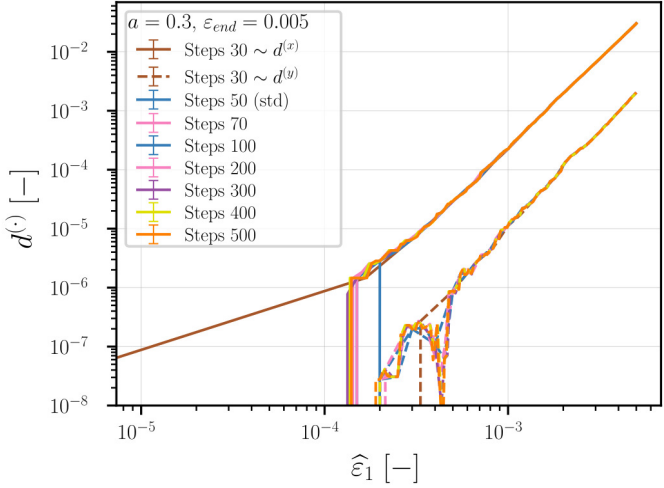

Figure C.18: Evolution of the damage eigenvalues $d^{(x)}$ (solid line) and $d^{(y)}$ (dashed line) for distortion level $a = 0.3$. The specimen was strained along the $x$-direction up to $\varepsilon_{end} = 0.005$. Different colours indicate number of load steps that were performed to reach it. For all cases the standard deviation (omitted) is roughly the same, see Fig. 7.

# References

Anciaux, G., O. Coulaud, J. Roman, and G. Zerah (2008). "Ghost force reduction and spectral analysis of the 1 D bridging method". In: *Bulletin of Sociological Methodology/Bulletin de Méthodologie Sociologique*. Vol. 37. doi: `10.1177/075910639203700105`.

Bauman, P. T., H. B. Dhia, N. Elkhodja, J. T. Oden, and S. Prudhomme (2008). "On the application of the Arlequin method to the coupling of particle and continuum models". In: *Computational Mechanics* 42.4. doi: `10.1007/s00466-008-0291-1`.

Bitencourt, L. A. G., O. L. Manzoli, P. G. C. Prazeres, E. A. Rodrigues, and T. N. Bittencourt (2015). "A coupling technique for non-matching finite element meshes". In: *Computer Methods in Applied Mechanics and Engineering* 290. doi: `10.1016/j.cma.2015.02.025`.

Brancherie, D. and A. Ibrahimbegovic (2009). "Novel anisotropic continuumdiscrete damage model capable of representing localized failure of massive structures: Part I: theoretical formulation and numerical implementation". In: *Engineering Computations* 26.1. Ed. by A. Ibrahimbegovic, I. Koar, and P. Marovi. doi: `10.1108/02644400910924825`.

Braun, M., I. Iváñez, and M. P. Ariza (2021). "A numerical study of progressive damage in unidirectional composite materials using a 2D lattice model". In: *Engineering Fracture Mechanics* 249. doi: `10.1016/j.engfracmech.2021.107767`.

Budarapu, P. R. and T. Rabczuk (2017). "Multiscale Methods for Fracture: A Review★". In: *Journal of the Indian Institute of Science* 97.3. doi: `10.1007/s41745-017-0041-5`.

Chen, H., Y. Jiao, and Y. Liu (2016). "A nonlocal lattice particle model for fracture simulation of anisotropic materials". In: *Composites Part B: Engineering* 90. doi: `10.1016/j.compositesb.2015.12.028`.

Chen, L., W. Sun, B. Chen, Z. Shi, J. Lai, and J. Feng (2021). "Multiscale study of fibre orientation effect on pullout and tensile behavior of steel fibre reinforced concrete". In: *Construction and Building Materials* 283. doi: `10.1016/j.conbuildmat.2021.122506`.

Chen, P. Y., M. Chantharayukhonthorn, Y. Yue, E. Grinspun, and K. Kamrin (2021). "Hybrid discrete-continuum modeling of shear localization in granular media". In: *Journal of the Mechanics and Physics of Solids* 153. doi: `10.1016/j.jmps.2021.104404`.

Delaplace, A. and R. Desmorat (2008). "Discrete 3D model as complimentary numerical testing for anisotropic damage". In: *International Journal of Fracture* 148.2. doi: `10.1007/s10704-008-9183-9`.

Desmorat, R., F. Gatuingt, and F. Ragueneau (2007). "Nonlocal anisotropic damage model and related computational aspects for quasi-brittle materials". In: *Engineering Fracture Mechanics* 74.10. doi: `10.1016/j.engfracmech.2006.09.012`.

Eliá, J. and G. Cusatis (2022). "Homogenization of discrete mesoscale model of concrete for coupled mass transport and mechanics by asymptotic expansion". In: *Journal of the Mechanics and Physics of Solids* 167. doi: `10.1016/j.jmps.2022.105010`.

Evangelista, F., G. d. S. Alves, J. F. A. Moreira, and G. O. F. d. Paiva (2020). "A globallocal strategy with the generalized finite element framework for continuum damage models". In: *Computer Methods in Applied Mechanics and Engineering* 363. doi: `10.1016/j.cma.2020.112888`.

Evangelista, F. and J. F. A. Moreira (2020). "A novel continuum damage model to simulate quasi-brittle failure in mode I and mixed-mode conditions using a continuous or a continuous-discontinuous strategy". In: *Theoretical and Applied Fracture Mechanics* 109. doi: `10.1016/j.tafmec.2020.102745`.

Farhat, C. and F.-X. Roux (1991). "A method of finite element tearing and interconnecting and its parallel solution algorithm". In: *International Journal for Numerical Methods in Engineering* 32.6. doi: `10.1002/nme.1620320604`.

Gaede, O., A. Karrech, and K. Regenauer-Lieb (2013). "Anisotropic damage mechanics as a novel approach to improve pre- and post-failure borehole stability analysis". In: *Geophysical Journal International* 193.3. doi: `10.1093/gji/ggt045`.

Guidault, P.-A. and T. Belytschko (2007). "On the L2 and the H1 couplings for an overlapping domain decomposition method using Lagrange multipliers". In: *International Journal for Numerical Methods in Engineering* 70.3. doi: `10.1002/nme.1882`.

Herrmann, H. J., A. Hansen, and S. Roux (1989). "Fracture of disordered, elastic lattices in two dimensions". In: *Physical Review B* 39.1. doi: `10.1103/PhysRevB.39.637`.

Lemaître, J., ed. (2001). *Handbook of materials behavior models*. San Diego, CA: Academic Press. 3 pp. isbn: 978-0-12-443341-0.

Lemaître, J. (1996). *A Course on Damage Mechanics*. Berlin, Heidelberg: Springer Berlin Heidelberg. doi: `10.1007/978-3-642-18255-6`.

Lemaître, J. and R. Desmorat (2005). *Engineering damage mechanics: ductile, creep, fatigue and brittle failures*. Berlin ; New York: Springer. 380 pp. doi: `10.1007/b138882`.

Liu, J. (2018). "Multi-scale FEM-DEM model for granular materials : micro-scale boundary conditions, statics and dynamics". PhD thesis. Technische Universiteit Eindhoven. isbn: 978-90-386-4621-3.

Lloberas-Valls, O., D. Rixen, A. Simone, and L. Sluys (2012a). "On micro-to-macro connections in domain decomposition multiscale methods". In: *Computer Methods in Applied Mechanics and Engineering* 225-228. doi: `10.1016/j.cma.2012.03.022`.

Lloberas-Valls, O., D. Rixen, A. Simone, and L. Sluys (2012b). "Multiscale domain decomposition analysis of quasi-brittle heterogeneous materials". In: *International Journal for Numerical Methods in Engineering* 89.11. doi: `10.1002/nme.3286`.

Matou, K., M. G. D. Geers, V. G. Kouznetsova, and A. Gillman (2017). "A review of predictive nonlinear theories for multiscale modeling of heterogeneous materials". In: *Journal of Computational Physics* 330. doi: `10.1016/j.jcp.2016.10.070`.

Mazars, J. and J. Lemaître (1985). "Application of Continuous Damage Mechanics to Strain and Fracture Behavior of Concrete". In: *Application of Fracture Mechanics to Cementitious Composites*. Ed. by S. P. Shah. Vol. 94. NATO ASI Series. Dordrecht: Springer Netherlands. doi: `10.1007/978-94-009-5121-1_17`.

Mier, J. G. M. van (2017). *Fracture Processes of Concrete: Assessment of Material Parameters for Fracture Models*. 1st ed. CRC Press. doi: `10.1201/b22384`.

Miller, R. E. and E. B. Tadmor (2009). "A unified framework and performance benchmark of fourteen multiscale atomistic/continuum coupling methods". In: *Modelling and Simulation in Materials Science and Engineering* 17.5. doi: `10.1088/0965-0393/17/5/053001`.

Moukarzel, C. and H. J. Herrmann (1992). "A vectorizable random lattice". In: *Journal of Statistical Physics* 68.5. doi: `10.1007/BF01048880`.

Oliver-Leblond, C., R. Desmorat, and B. Kolev (2021). "Continuous anisotropic damage as a twin modelling of discrete bidimensional fracture". In: *European Journal of Mechanics - A/Solids* 89. doi: `10.1016/j.euromechsol.2021.104285`.

Ostoja-Starzewski, M. (2008). *Microstructural randomness and scaling in mechanics of materials*. CRC series–modern mechanics and mathematics. Boca Raton: Chapman & Hall/CRC. 471 pp. doi: `10.1201/9781420010275`.

Reddy, J. N. (1997). "On locking-free shear deformable beam finite elements". In: *Computer Methods in Applied Mechanics and Engineering*. Containing papers presented at the Symposium on Advances in Computational Mechanics 149.1. doi: `10.1016/S0045-7825(97)00075-3`.

Reddy, J. N., C. M. Wang, and K. Y. Lam (1997). "Unified Finite Elements Based on the Classical and Shear Deformation Theories of Beams and Axisymmetric Circular Plates". In: *Communications in Numerical Methods in Engineering* 13.6. doi: `10.1002/(SICI)1099-0887(199706)13:6<495::AID-CNM82>3.0.CO;2-9`.

Rezakhani, R., X. Zhou, and G. Cusatis (2017). "Adaptive multiscale homogenization of the lattice discrete particle model for the analysis of damage and fracture in concrete". In: *International Journal of Solids and Structures* 125. doi: `10.1016/j.ijsolstr.2017.07.016`.

Richart, N. and J. F. Molinari (2015). "Implementation of a parallel finite-element library: Test case on a non-local continuum damage model". In: *Finite Elements in Analysis and Design* 100. doi: `10.1016/j.finel.2015.02.003`.

Rodrigues, E. A., O. L. Manzoli, L. A. G. Bitencourt, T. N. Bittencourt, and M. Sánchez (2018). "An adaptive concurrent multiscale model for concrete based on coupling finite elements". In: *Computer Methods in Applied Mechanics and Engineering* 328. doi: `10.1016/j.cma.2017.08.048`.

Sun, B. and Z. Li (2015). "Adaptive concurrent multi-scale FEM for trans-scale damage evolution in heterogeneous concrete". In: *Computational Materials Science* 99. doi: `10.1016/j.commatsci.2014.12.033`.

Sun, X. and X. Guo (2019). "Domain information transfer method and its application in quasi-brittle failure analysis". In: *Advances in Mechanical Engineering* 11.12. doi: `10.1177/1687814019895736`.

Unger, J. F., S. Eckardt, and C. Konke (2011). "A mesoscale model for concrete to simulate mechanical failure". In: *Computers and Concrete* 8.4. doi: `10.12989/CAC.2011.8.4.401`.

Unger, J. F. and S. Eckardt (2011). "Multiscale Modeling of Concrete". In: *Archives of Computational Methods in Engineering* 18.3. doi: `10.1007/s11831-011-9063-8`.

Vardoulakis, I. (2019). *Cosserat Continuum Mechanics: With Applications to Granular Media*. Vol. 87. Lecture Notes in Applied and Computational Mechanics. Cham: Springer International Publishing. doi: `10.1007/978-3-319-95156-0`.

Wellmann, C. and P. Wriggers (2012). "A two-scale model of granular materials". In: *Computer Methods in Applied Mechanics and Engineering*. Special Issue on Advances in Computational Methods in Contact Mechanics 205-208. doi: `10.1016/j.cma.2010.12.023`.

Wittel, F. K. (2006). "Diskrete Elemente-Modelle zur Bestimmung der Festigkeitsevolution in Verbundwerkstoffen". PhD Thesis. Universität Stuttgart. isbn: 3-930683-59-8.

Xiao, S. P. and T. Belytschko (2004). "A bridging domain method for coupling continua with molecular dynamics". In: *Computer Methods in Applied Mechanics and Engineering*. Multiple Scale Methods for Nanoscale Mechanics and Materials 193.17. doi: `10.1016/j.cma.2003.12.053`.

Xu, L., L. Jiang, L. Shen, L. Gan, Y. Dong, and C. Su (2023). "Adaptive hierarchical multiscale modeling for concrete trans-scale damage evolution". In: *International Journal of Mechanical Sciences* 241. doi: 10.1016/j.ijmecsci.2022.107955.

Zhang, H., J. Wu, and Y. Zheng (2012). "An adaptive multiscale method for strain localization analysis of 2D periodic lattice truss materials". In: *Philosophical Magazine* 92.28. doi: 10.1080/14786435.2012.731087.